\documentclass[aps,pre,twocolumn,floatfix,showpacs]{revtex4}
\usepackage{amssymb}

\usepackage{amsmath}
\usepackage{graphicx}

\begin{document}

\preprint{}
\title[ShortTitle]{ Potential landscapes and induced charges near
metallic islands in three dimensions}
\author{A.~Weichselbaum}
\author{S.~E.~Ulloa}
\affiliation{Department of Physics and Astronomy, and Nanoscale
and Quantum Phenomena Institute, Ohio University, Athens, Ohio
45701}
 \pacs{02.70.-c, 41.20.-q}

\begin{abstract}
We calculate electrostatic potential landscapes for an external
probe charge in the presence of a set of metallic islands.  Our
numerical calculation in three dimensions (3D) uses an efficient
grid relaxation technique. The well-known relaxation algorithm for
solving the Poisson equation in two dimensions is generalized to
3D. In addition, all charges on the system, free as well as
induced charges, are determined accurately and self-consistently
to satisfy the desired boundary conditions. This allows the
straightforward calculation of the potential on the outer boundary
using the free space electrostatic Green's function, as well as
the calculation of the entire capacitance matrix of the system.
Physically interesting examples of nanoscale systems are presented
and analyzed.
\end{abstract} 

\volumeyear{year}
\volumenumber{number}
\issuenumber{number}
\eid{identifier}
\date[Date: ]{June 20, 2003}
\maketitle


\section*{Introduction}

There is a need to precisely know the electrostatic landscape
experienced by electrons in ever smaller structures, down to the
scale of scanning tunnelling microprobes and single electron
devices. The presence of conducting leads for manipulating and
measuring local potentials influences the quantum mechanical
behavior of electrons in a highly non-trivial manner. In
polarizable media, for example, charged ``conglomerates" which
include free as well as polarization charges in the neighborhood,
behave as quasi-particles which can live for a comparatively long
time and interact with the conducting leads via Coulomb
interaction. Knowledge of the potential landscape describing this
interaction for the case of a quasi-particle, is an interesting
and important element in the better understanding of these
systems.  Electrons in single-electron transistors \cite{Likh99},
or moving in the neighborhood of lithographically-defined gate
arrangements \cite{2DEG-ref}, or tunnelling through STM scanning
tips, are but a few examples of the pervasiveness of electrostatic
potentials in realistic structures.

The solution of the Laplace or Poisson equation to obtain
electrostatic potential landscapes is a well defined boundary
value problem, typically requiring the discretization of space on
a convenient grid.  Relaxation techniques are well known
\cite{NumRe93, Demmel97, Jack99} and widely used in the solution
of these problems, as they provide convenient and efficient
algorithms \cite{coalson98}. For dimensions lower than three,
simple second order algorithms are used together with common
speedup features such as successive overrelaxation (SOR) and
Gauss-Seidel (GS) iteration schemes \cite{NumRe93, Demmel97}. In
three dimensions, however, due to the poor scaling with grid
dimensions, more efficient routines are desirable. In this context
a generalized $O(h^{6})$ algorithm for 3D is presented in this
paper, and used to calculate the potential landscape of several
physical systems of interest, as those mentioned above.

The boundary conditions most easily built into relaxation
algorithms are fixed voltage surfaces, with the voltage known and
provided by an external battery, for example. This does not apply,
however, to cases with a {\em floating} potential, such as
metallic islands which are isolated from the environment (like
metallic quantum dots), or for open boundary problems.  In the
case of isolated islands, the value of the potential at a metallic
boundary, even though constant, is not known. On the other hand,
the overall charge on the island is determined at the outset and
can be considered to be known. The solution to the problem taken
here is that once one has access to the linear relationship
between the charge on the island and the island potential derived
from the relaxation algorithm, one can invert this relationship
and calculate the potential with every relaxation cycle such
that the overall charge is maintained at a fixed value,
e.g. zero for an overall neutral island.
Incorporation of this idea in the iteration procedure yields the
appropriate floating potentials, as we will show.

Notice moreover that the outer boundary is open in general in
most nano-sized geometries. If the
size of the calculated cell could be chosen large enough, of
course, one could assume that the potential would drop to zero
there; however, this is operationally forbidden by the vast number
of grid points needed in that case. The only feasible way is to
determine the non-uniform potential on the outer boundary
self-consistently within the algorithm. The approach taken in this
paper is that the knowledge of the total charge distribution
(external \textit{and} induced charges) allows for the calculation
of the potential on the outer boundary via the standard free-space
electrostatic Green's function, $1/4\pi r $. This is a
straightforward if computationally expensive procedure, which can
be speeded up remarkably by tabulating the inverse distances on
the grid, but yields physically well-behaved asymptotics in all
cases, as we will see.  We should also emphasize that once the
charges and potentials in the system are known, one can easily
evaluate the capacitance matrix for the geometries of interest,
regardless of the symmetries of the arrangement.

The remainder of the paper describes the algorithm in detail in
section I, while section II illustrates its use in several
physically relevant examples.

\section{Algorithm}

Taking the standard Taylor expansion of a smooth function
$f(x,y,z)$ around each grid point, one defines the \textit{center
average} and the \textit{square average} as follows (assuming
uniform grid spacing $h$ in all three dimensions)
\begin{subequations}
\label{avg3D}
\begin{eqnarray}
\left\langle f\right\rangle _{C} &\equiv &\frac{1}{6}\left(
\sum_{i,j,k}^{NN}f_{ijk}\right) =  \notag \\
&=&f^{(000)}+\frac{h^{2}}{6}\vec{\nabla}^{2}f+\frac{h^{4}}{72}\left(
f^{(400)}+f^{(040)}+f^{(004)}\right)  \notag \\
&&+O\left( h^{6}\right)  \label{avg3D_C} \\
\left\langle f\right\rangle _{S} &\equiv &\frac{1}{8}\left(
\sum_{i,j,k}^{TNN}f_{ijk}\right) =  \notag \\
&=&f^{(000)}+\frac{h^{2}}{2}\vec{\nabla}^{2}f+\frac{h^{4}}{24}\left(
f^{(400)}+f^{(040)}+f^{(004)}\right)  \notag \\
&&+\frac{h^{4}}{4}\left( f^{(220)}+f^{(202)}+f^{(022)}\right)
+O\left( h^{6}\right) \, ,  \label{avg3D_S}
\end{eqnarray}%
\newline
where $(T)NN$ stands for (third) nearest neighbors and
\end{subequations}
\begin{subequations}
\label{defs_fg}
\begin{eqnarray}
&&f^{(rst)}\equiv \frac{\partial ^{r}}{\partial x^{r}}\frac{\partial ^{s}}{%
\partial y^{s}}\frac{\partial ^{t}}{\partial z^{t}}f\left( i,j,k\right)
\label{def_deriv} \\
&&f_{\mathbf{i}}\equiv f_{ijk}\equiv f(i,j,k)\equiv f(i\cdot
h,j\cdot h,k\cdot h) \, , \label{def_gpos}
\end{eqnarray}%
with $i,\,j,\,k$ and $r,\,s,\,t$ being integers. The averages in
Eqs.\ (\ref{avg3D}) are shown graphically in Fig.\
\ref{fig_avg_cs}. Notice that the odd-order derivatives in
eqs.~(\ref{avg3D}) cancel due to the symmetric combinations around
the grid points included in the averages.  Notice also that {\em
second} (or next) nearest neighbors are considered in the ``checkered
lattice" sweeps of the points making the simple cubic grid. (The
relaxation sweeps are done sequentially over the face-centered cubic
array of neighbors which form effectively a dual lattice.) 

\begin{figure}[tbp]
\includegraphics*[width=8.6cm]{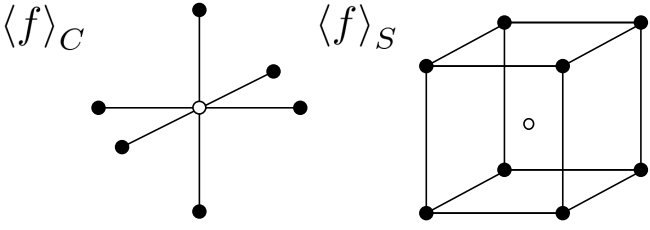}
\caption{Visualization of averages taken on the grid: average
$\left\langle f\right\rangle _{C}$ as in Eq.\ (\ref{avg3D_C}) is
over nearest neighbors (NN); average $\left\langle f\right\rangle%
_{S}$ as in Eq.\ (\ref{avg3D_S}) is over third nearest neighbors
(TNN). The grid points included in the sum are shown as filled
symbols; solid lines joining these have length of two grid
spacings, $2h$.} \label{fig_avg_cs}
\end{figure}

Taking the linear combination of the averages (\ref{avg3D_C}) and
(\ref {avg3D_S})
\end{subequations}
\begin{equation}
\left\langle \left\langle f\right\rangle \right\rangle \equiv
\alpha \,\left\langle f\right\rangle _{C}+\left( 1-\alpha \right)
\left\langle f\right\rangle _{S} \, ,  \label{avg3D_alpha}
\end{equation}%
then with $\alpha \equiv \alpha _{3D}=\frac{6}{7}$ the \textit{overall
average} becomes
\begin{eqnarray}
\left\langle \left\langle f\right\rangle \right\rangle _{3D} &=&f_{\mathbf{i}%
}+\frac{3}{14}h^{2}\,\vec{\nabla}^{2}f_{\mathbf{i}}+\frac{1}{56}h^{4}\,\,%
\vec{\nabla}^{2}\,\vec{\nabla}^{2}f_{\mathbf{i}}  \notag \\
&&+O\left( h^{6}\right) \, . \label{avg3D_all}
\end{eqnarray}%
Since we are solving for the Poisson equation%
\begin{equation}
\,\vec{\nabla}^{2}f=-g \, , \label{poisson_eq}
\end{equation}%
Eq.\ (\ref{avg3D_all}) can be rewritten as%
\begin{eqnarray}
f_{\mathbf{i}} &=&\left\langle \left\langle f\right\rangle \right\rangle
_{3D}+\frac{3}{28}h^{2}\,\left( g_{\mathbf{i}}+\left\langle g\right\rangle
_{C}\right)  \notag \\
&&+O\left( h^{6}\right) \, , \label{relax_eq}
\end{eqnarray}%
where $\left\langle g\right\rangle _{C}$ is the \textit{center
average} for the \textit{source term} at a given grid point.
Equation~(\ref{relax_eq}), together with (\ref{avg3D_all}), serves
as the basis for the iterative scheme: the potential $f$ is
\textit{relaxed to its minimum} considering external sources $g$
and the potential of the (third and) nearest neighbor sites
through the linear 3D average. In order to calculate charges on
the grid, Eq.\ (\ref{avg3D_all}) is employed again and gives a
straightforward way for calculating \textit{real} (i.e. free) and \textit{%
induced} charges on the grid on an equal footing. Thus using Poisson's
equation (\ref{poisson_eq}), the charge contribution of each grid point can
be calculated as follows%
\begin{eqnarray}
q_{\mathbf{i}} &\equiv&h^{3}\,g_{\mathbf{i}}=h\,\left( -h^{2}\vec{\nabla}^{2}f_{%
\mathbf{i}}\right) =  \notag \\
&=&\frac{14}{3}h\,\left( f_{\mathbf{i}}-\left\langle \left\langle
f\right\rangle \right\rangle _{3D}+O\left( h^{4}\right) \right) \,
. \label{eq_qinduced}
\end{eqnarray}

For convenience, throughout the paper we adopt the following
convention on units: The charge of an electron is taken to be $1$
and the Coulomb energy is written
in units of ${\rm eV}$, as $E=q_{1}\left[ e\right] \,\cdot \,q_{2}\left[ e%
\right] \,/\,(4\pi r)$; this straightforwardly implies a unit for
the distance of $[r]=18.1$ nm.  In short, the units chosen are
\begin{equation}
\lbrack
energy]=\rm{eV};\;[charge]=\mathrm{e};\;[distance]=18.1\rm{nm}
\label{eq_units}
\end{equation}%
Therefore taking these units, brings one naturally into the realm
of nanostructures.

\subsection{Successive over-relaxation and iteration scheme}

The general method for successive over-relaxation (SOR) is
described for 2D in \cite{NumRe93, Demmel97} for a $N\times
N$ array, 
and generalized here to 3D as given by
\begin{subequations}
\begin{eqnarray}
f_{\mathbf{i}}^{(i+1)} &=&f_{\mathbf{i}}^{(i)}+\omega \,\left( f_{\mathbf{i}%
}^{(new)}-f_{\mathbf{i}}^{(i)}\right) \text{,}  \label{eq_sor} \\
\text{where } \notag \\ \omega &=&\frac{2}{1+\pi /\min
(N_{x},N_{y},N_{z})} \, , \label{eq_sor_om}
\end{eqnarray}%
\newline
$N_j$ is the grid size in the $j$th direction, and
$f_{\mathbf{i}}^{(new)}$ is calculated according to Eq.\ (\ref%
{relax_eq}). The SOR parameter $\omega $ is in the range $1<\omega
<2$ as required for the algorithm to converge. The basic idea
behind $\omega $ is that if one is heading in the \textit{right}
direction (e.g. towards the solution), why not go a bit further.
An $\omega $ too large ($\omega >2$), however, results in
instability of the algorithm and the relaxation process overshoots
and diverges. Equation~(\ref{eq_sor_om}) was tested for different
$N_{x},\,N_{y}$ and $N_{z}$ values and it was indeed this specific
combination that gave the optimal value for $\omega $ (note that
the window for a \textit{good} $\omega $ is quite narrow in
general). The specific structure of (\ref{eq_sor_om}) can be
intuitively understood as follows: the SOR algorithm introduces
perturbations in the system that propagate during the relaxation
cycles and eventually die out if the grid is large enough;
however, for finite grid sizes, the perturbations are reflected at
the boundaries and so they interfere and pile up. In this sense,
the minimum extension within the three grid dimensions constrains
the optimal magnitude of $\omega $, consistent with Eq.\
(\ref{eq_sor_om}).

For further optimization of the algorithm, the Gauss-Seidel
iteration scheme was adopted, as well as the alternating
relaxation on the two checkerboard like sub-grids that in sum span
the whole grid \cite{NumRe93, Demmel97}. The \textit{inverse}
distances between the grid points were mapped into a table, such
that the calculation of the potential in the grid is sped up
remarkably. According to Coalson \cite{coalson98}, multigrid
methods can be applied to account for the slowly converging long
wavelength portions of the solution. This was not done here, since
the variation of the potential on the isolated islands and
especially the calculation of the outer boundary already
introduced longer range correlations over the grid that presumably
made the algorithm converge faster in our case.

A note about efficiency:  As we use a successive overrelaxation
method to iterate the potential on the grid, the total relaxation
time for this in 2D is proportional to $\sim n^{3/2}$, where $n$
is the {\em total} number of grid points \cite{Demmel97}, and is
thus clearly comparable to algorithms like conjugate gradient.
Notice also that SOR has still known improvements that may also be
implemented and would thus make this algorithm superior to the
former \cite{Demmel97}. Our relaxation over the bipartite lattices
composing the simple cubic 3D grid preserves the spirit of the 2D
algorithms, but obtains an accuracy of $O(h^{6})$, as discussed
above.

\subsection{Open outer boundary}

Equation (\ref{eq_qinduced}) gives a consistent higher order
recipe for calculating the overall charge distribution, including
induced as well as the external (free) charges
\cite{remark_calc_qi, remark_calc_qi_2}, given via the source term
$g$ in the Poisson equation (\ref{poisson_eq}). Starting with an
(arbitrary) initial constant potential on the outer boundary (OB),
these values are updated every time the interior of the grid has
relaxed down to a certain accuracy level $\varepsilon $,
\end{subequations}
\begin{equation*}
V_{\mathbf{i}}^{OB}=\frac{1}{4\pi }\sum_{q_{\mathbf{j}} \neq
0}q_{\mathbf{j} }\cdot \left( d^{-1}\right) _{\mathbf{ij}}\; ,
\end{equation*}%
and employs the tabulated inverse distance values for the grid
points.

\section{Discussion}

In the following, several applications of the algorithm are
presented. We start with a simple test example, and follow with
the analysis of more complex geometries.

\subsection{Example: point charge near a conducting island}

\begin{figure}[tbp]
\includegraphics*[width=8.6cm]{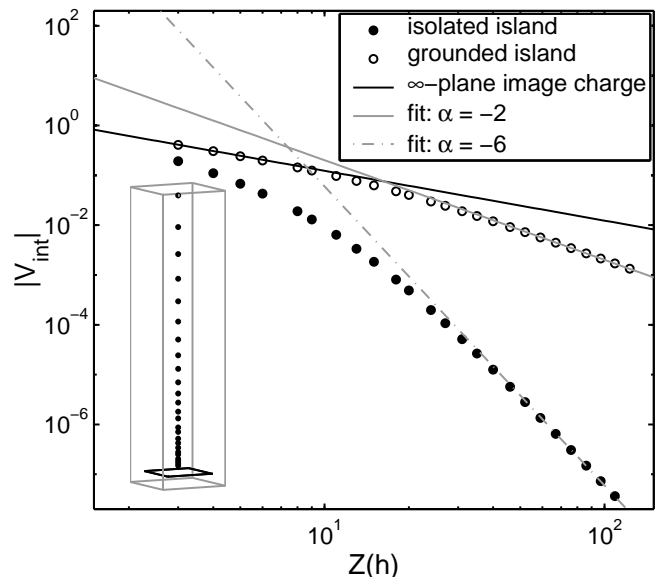}
\caption{Calculated interaction potential of a point charge
($q=1\left[ e\right] $) with a finite size conducting plane. The
grid configuration is shown to scale in the inset: the total block
size is 32$\times $32$\times $132, indicated by the gray box
lines; positions of the point charge are indicated by the black
dots in the inset, a set of locations vertically away from the
metallic plane which is shown at the bottom by the black line
(note that closest point-charge to island distance is $3h$, three
grid spacings). Shown are two configurations: neutral isolated
island (filled circles) and grounded island (empty circles). Gray
lines indicate different power law dependencies $\varpropto z^{%
\protect\alpha }$.} \label{fig_Vint_isol_grnd}
\end{figure}

As an instructive example and as a test case for the algorithm, a
$32\times 32\times 132$ grid was setup with one square metallic
island in the lower region, while an external charge is placed at
different positions away from the island surface and directly over
its center (see inset of Fig.\ \ref{fig_Vint_isol_grnd}). The
total dimensions of the grid in real space were taken to be
$L_{x}=1$ (in units as per \ref{eq_units}), and accordingly
$L_{y}=1$ and $L_{z}=4.2$ for equal grid spacing.

We calculate the interaction potential experienced by the point
charge in the presence of the island. The results are shown in
Fig.\ \ref {fig_Vint_isol_grnd}. For the case of a neutral island,
close to the surface, the potential approaches that produced by
the image charge of an infinite plane ($ \sim 1/r$), while further
away from the island, the potential approaches asymptotically the
form of an induced quadrupole interaction potential ($ \sim 1/r^{6}$)
\cite{remark_on_r6}.

The calculation was done also for a grounded island. In this case,
one obtains the limit of an infinite plane at short distances
($\sim 1/r$), while further away from the island the dependence
weakens to $ \sim 1/r^{2}$ as expected  \cite{remark_on_r2}.
As a reference, the interaction potential in the presence of an
infinite plane is shown (solid line in Fig.\ \ref {fig_Vint_isol_grnd}).
The finite size of the island clearly reduces the interaction at
large distances. Overall, it is interesting to see
that one obtains exactly what is
expected, but more importantly, the algorithm already gives
useful results for rather small grid dimensions.

\subsection{Example: point charge interacting with an array of islands}

\begin{figure}[tbp]
\includegraphics*[width=8.6cm]{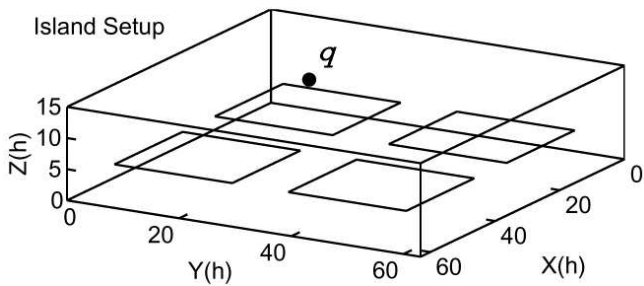}
\caption{Interacting point charge ($q=+1$ ) with an array of four
islands in a plane underneath the point charge: a $64\times
64\times 16$ (box) cell
with open boundaries was setup; its dimensions are $L_{x}=%
L_{y}=1$, $L_{z}=16h=0.25$ (units as per (\ref{eq_units})).  The
separation of the external charge from the island plane is
$5h=5/64=0.078$.} \label{fig_4isl_setup}
\end{figure}

Lithographically, an array of conducting islands can be separated
from a 2D electron gas by an insulating layer, as in the
experiments with spatially modulated two-dimensional electron
gases in semiconductors \cite{2DEG-ref}. Considering the electron
gas as a Fermi liquid, the interaction of a single electron (or
single quasi particle) with the conducting islands nearby is an
important element of the physics of this problem since this
interaction will clearly modify the dynamical behavior of the
system \cite{Likh99}. Figure~\ref{fig_4isl_setup} shows the sample
geometry used in this example.

\begin{figure}[tbp]
\includegraphics*[width=8.6cm]{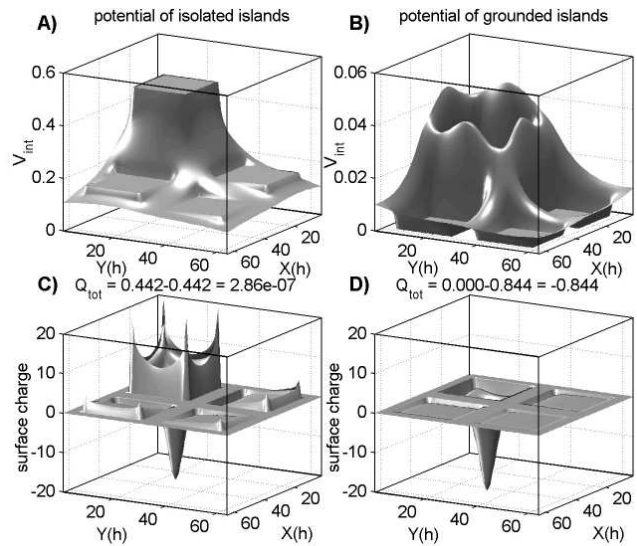}
\caption{ Potential landscape and charge distribution
\textit{within} the plane of islands for the geometry in Fig.\
\ref{fig_4isl_setup}: the left (right) two graphs are for the case
of neutral (grounded) islands, respectively. Notice the quite
different scale in the two potentials shown in A and B. The
induced charge is shown as surface charge in the plots C and D
where for better visualization the top (positive values) and
bottom surface (negative) charges of the islands are shown
together in a single plot.} \label{fig_4isl_vq}
\end{figure}

\begin{figure}[tbp]
\includegraphics*[width=8.6cm]{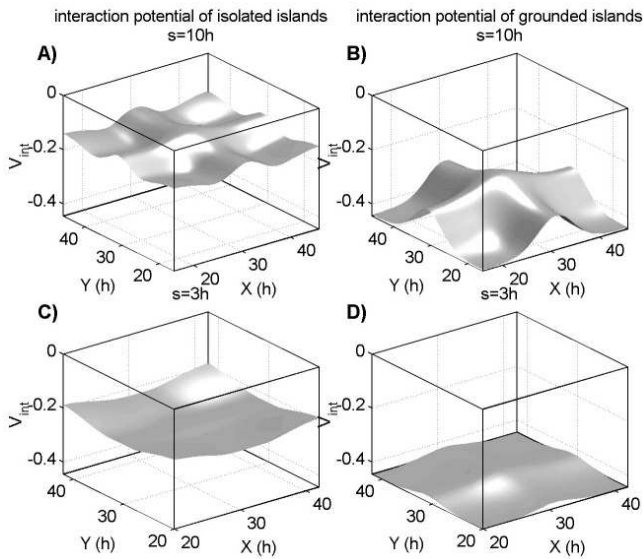}
\caption{Potential landscape of an external charge interacting
with the array of islands. The geometry is the same as in Fig.\
\ref{fig_4isl_setup} except for the separation between islands.
Only the region of the four overlapping island corners is shown.
The left two graphs, A and C, are for the case of isolated
islands; the right two, B and D, for grounded islands; the upper
two graphs are for an island separation $s$ of $10h$ ($= 10$ grid
spacings), while the lower two graphs are for an island separation
of $3h$.} \label{fig_4isl_vint}
\end{figure}

With the test charge hovering over the center of one island and
considering the conducting islands either grounded or isolated
(uncharged), the calculated charge distribution and the potential
in the plane of the islands are shown in Fig.\ \ref{fig_4isl_vq};
the grounded case can be understood as that of an island where
hopping onto and off the island is allowed via a tunnelling
channel from an external charge reservoir. A few points should be
stressed: first, note that the variations of the potential in the
plane of islands in Fig.\ \ref{fig_4isl_vq} for the case of
isolated islands is about a factor $10$ {\em larger} compared to
the grounded case since the island potential is not fixed. Second,
the isolated islands are indeed neutral within numerical accuracy
(the sum over all charges $Q_{tot}$ is zero, within the accuracy
provided by the convergence parameter $\varepsilon $ (see lower
two plots in Fig.\ \ref {fig_4isl_vq}: total induced charge in the
plane of the islands = all positive surface charge + all negative
surface charge $\simeq 0.442-0.442\sim 10^{-7} \sim \varepsilon%
$). Third, the induced negative charge in the case of neutral
islands is clearly smaller compared to the grounded case, which is
intuitively clear, since in the neutral case the negative charge
needs to be compensated by an equal amount of opposite charge, and
this charge separation in the island costs energy. Therefore, the
interaction with the external charge can be expected to be weaker
for the isolated (neutral) islands, as is the case (see later,
Fig.\ \ref{fig_4isl_vint}). Notice also that in the neutral case
the island corners exhibit accumulation of induced charges, as one
would expect from the sharpness of the island corners. It is also
interesting to observe that the induced charge never exceeds the
external charge (in absolute value): at the maximum, it is just
equal and opposite in sign (as for the case of the infinite
plane).

As the external charge was displaced horizontally at a certain
separation above the plane of the island, one effectively scans
the potential landscapes for different geometries and different
boundary conditions, as shown in Fig.\ \ref{fig_4isl_vint}. The
isolated (neutral) islands show a clearly weaker modulation of the
interaction potential. Furthermore, decreasing the island
separation such that the gap between them is nearly closed,
reduces most of the central modulation and smooths the potentials,
as one would expect, and can be seen by comparing the upper two
graphs in Fig.\ \ref{fig_4isl_vint} with the lower two.

\subsection{Mutually induced charge on STM tip}

\begin{figure}[t]
\includegraphics*[width=8.6cm]{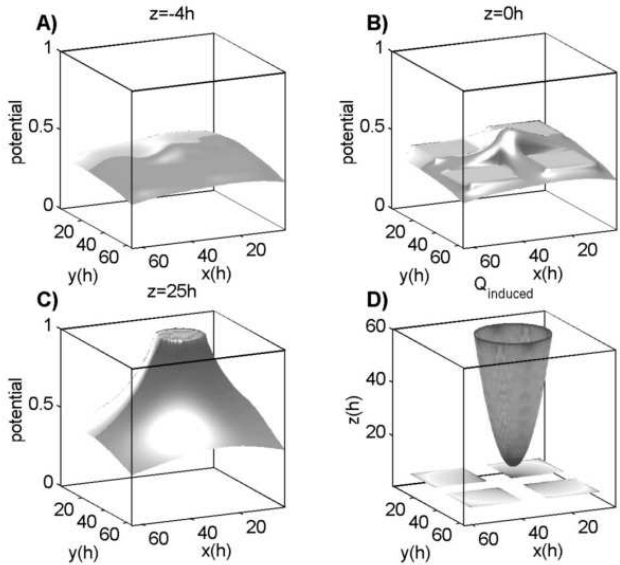}
\caption{Potential and charge distribution for a conducting tip at
potential $V=1$ above four isolated (neutral) islands; the
distance between the plane of islands and the tip is $5h$. Panels
A to C show $xy$-slices in the potential distribution. Panel A:
potential just under the plane of islands ($z=-4h$ with respect to
the plane of the islands); panel B: potential right on the plane
of islands ($z=0h$); and panel C: potential on a plane above the
tip ($z=25h$). Panel D is a contour plot of the charge
distribution on the tip and on the overall neutral islands
($ Q_{tot}^{islands}=+0.281+(-0.281)\simeq 10^{-10}$, i.e. nicely converged).
The {\em extra} charge on the tip due to the presence of the
islands is $+0.181$.} \label{fig_stip_vq}
\end{figure}

As a final example, the induced charges where calculated for the
typical geometry of a metallic STM tip near a metallic array
structure with the islands kept neutral;
the geometry can be seen in panel D of Fig.\
\ref{fig_stip_vq}. The parabolically shaped STM tip is maintained
at a fixed potential ($V=1$) by an external source. The potential
below the islands is weakened and shielded by the islands (panel A),
while on the islands the potential is constant and is defined there by
the constraint of neutrality (panel B). Above the islands, in the region of
the tip, the potential landscape ($xy$-slice) has a circular
plateau at the position of the tip with a potential of $V=1$ (the
potential of the tip) and smoothly decays away from the tip (panel C).
Panel D indicates the charge distribution on the islands and on the tip;
the charge on the tip is positive overall: $ Q_{tip}^{0}=+4.01$ is
the charge on the portion of the tip shown in the \textit{absence}
of the islands, and it gains a bit more of charge ($+0.181$)
through the interaction with the metallic islands. The island with
the tip right on top of it (right rear in inset), is the most
polarized of all. The corner in the center is \textit{negatively}
charged, while the transition through the white region on the
island surfaces in panel D indicates that the surface charge
switches sign there, and that the outer region of the islands are
positively charged.  In addition, the lower island surfaces are
also mostly positive, as one would expect. This compensates the
negative charge induced by the tip and guarantees the neutrality
of the islands (see caption for explicit numbers).

\section{Summary}

In summary, the electrostatic potential of complex metallic
arrangements were calculated on a three dimensional grid with an
$O(h^{6})$ algorithm.  The algorithm presented here is a
generalization of the relaxation techniques common in 2D systems,
properly set up to provide accurate and efficient calculations.
The approach allows the study of arbitrary geometries and boundary
conditions, as well as the self-consistent calculation of free and
induced charges. This information, in turn, allows the calculation
of the capacitance matrix of the system. Several examples
illustrate the reliability and usefulness of the algorithm for
obtaining potential landscapes of interest.

\begin{acknowledgements}
We acknowledge support from NSF Grant NIRT 0103034, and the
Condensed Matter and Surface Sciences Program at Ohio University.
\end{acknowledgements}

\newif\ifabfull\abfulltrue


\begin{thebibliography}{1}
\bibitem{remark_calc_qi}
Calculating the external charges (e.g. that of an isolated
electron) is not needed explicitly since known, but it serves as
nice self-check of the algorithm. As an example, the charge
distribution around a free electron was calculated, introduced
into the system as a $\delta $-charge (i.e., a source of $1/h^{3}$
on a single grid point).  With nearest neighbor sites included
because of the discreteness of the lattice, the total charge over
these $1+6$ grid points gave correctly the overall charge up to
numerical precision.
\bibitem{remark_calc_qi_2}
When calculating an artificial charge at the outer boundary by
assuming continued constant potential right at the border results
in the constraint that the sum of charges from all grid points
needs to approach zero.  This was also nicely fulfilled up to a
relative error of $10^{-7}$, the order of the convergence
parameter  $\varepsilon$.
\bibitem{remark_on_r6}
The quadrupole dependence is obtained because the island is taken
ideally thin (one grid layer) and therefore, out of symmetry of
the chosen configuration, there is no dipole moment; with this,
the far field $\sim 1/r^6$ dependence for a quadruple
is obtained in a similar fashion as in \cite{remark_on_r2}.
\bibitem{remark_on_r2}
The far field $\sim 1/r^2$ dependence for a grounded conductor can
be obtained by minimizing the energy of the probe charge in the
presence of the charged conductor, $E = Q^2/2C + qQ/z$, where $q$
is the probe charge, $Q$ than on the conductor of capacitance $C$,
and $z$ is their separation.  The minimal energy yields $Q=-qC/z$,
so that the potential on $q$ is then $qQ/z = -q^2C/z^2$.
\end{thebibliography}
\end{document}